# Incorporating Gibbs free energy into interatomic potential fitting

Liangrui Wei and Yang Sun*

*Department of Physics, Xiamen University, Xiamen 361005, China*

(Dated: Jan. 12, 2026)

We develop a method to fit high-temperature Gibbs free energy data for the development of interatomic potentials for atomic systems. The approach is based on Hamiltonian thermodynamic integration, enabling the identification of suitable potential parameters such that the system's free energy matches a specified target. The method can be readily combined with conventional fitting techniques for properties such as elastic tensors and liquid pair distribution functions. We validate the effectiveness of the approach using the Uhlenbeck–Ford model and embedded-atom method potentials for pure Ni phases and binary $Fe_{1-x}O_x$ liquids under high-pressure and high-temperature conditions. Our framework provides an efficient strategy for incorporating free energy into interatomic potential fitting.

## I. INTRODUCTION

The success of molecular dynamics simulations in predicting materials' properties largely depends on how interatomic interactions are described. *Ab initio* molecular dynamics (AIMD) inherits the accuracy of first-principles electronic structure calculations based on density functional theory (DFT) [1–4], but is often limited by the size of the simulation cell and the accessible time scales. Classical molecular dynamics (CMD) allows for significantly larger and longer simulations but requires an accurate interatomic potential [5–12]. These potentials are typically designed to reproduce properties obtained from AIMD simulations or experimental data, enabling more detailed and large-scale modeling. Recent developments in machine-learning interatomic potentials primarily focus on reproducing atomic force and total energy from DFT [13–15]. In contrast, semi-empirical potentials, built on physically motivated functional forms, often have a direct connection to target physical properties such as the lattice constant, elastic tensor, and liquid structure [16–20].

The limited number of fitting parameters in semi-empirical potentials makes them computationally efficient, but sometimes makes it difficult to achieve satisfactory accuracy for properties of interest. In particular, directly fitting high-temperature properties, such as solidus and liquidus lines in multicomponent phase diagrams or the nucleation driving force under supercooled conditions, remains difficult [21]. Efforts have been made to fine-tune such high-temperature quantities via thermodynamic relations, for instance, using the Gibbs–Duhem equation to adjust the melting points of unary systems [22]. In principle, the Gibbs free energy is the most fundamental thermodynamic quantity for reproducing high-temperature behavior, as all thermodynamic properties can be derived from the Gibbs free energy, or its derivatives with respect to temperature, pressure, and chemical composition. Thus, directly fitting the Gibbs free energy is essential for developing accurate semi-empirical potentials for high-temperature applications.

High-temperature Gibbs free energy data is key to potential fitting. Enthalpy data at 0 K can be obtained with high accuracy via DFT calculations, making it a commonly used fitting target in potential development [23]. However, first-principles high-temperature free energy used to be hard to compute, resulting in a lack of fitting data in earlier potential development efforts. The rapid increase in computational power, along with algorithmic optimizations, has now enabled the direct calculation of Gibbs free energy from *ab initio* simulations, especially for materials under extreme conditions that are challenging to probe experimentally [24–29]. Deep learning techniques can also generate free energy data with accuracy close to *ab initio* level [30,31]. These free energy data thus offer precise fitting targets for the development of semi-empirical potentials.

In this work, we aim to develop a method to adjust the Gibbs free energy of a semi-empirical potential to match a target value. We propose employing Hamiltonian thermodynamic integration (HTI) in the fitting process. We demonstrate that HTI can be effectively combined with existing fitting techniques for properties such as the liquid pair correlation function, elastic constants, etc., enabling the development of accurate potentials capable of reproducing free energy obtained from *ab initio* calculations.

This paper is organized as follows. Section II derives the formulas for free energy fitting and provides details of the semi-empirical potential and the MD simulations used in this work. Section III evaluates the effectiveness of the method on a model system, as well as on single-component and binary systems. In Section IV, we discuss possible applications to other types of potentials. Finally, our conclusions are summarized in Section V.

---

*Email: yangsun@xmu.edu.cn

## II. METHODS
### A. Free energy fitting

We consider a classical atomic system with atom coordinates $\mathbf{R} = \{\mathbf{r}_i\}$, where $i$ runs for the atom index. $\mathbf{X} = (x_1, x_2, \ldots x_k)$ is the set of parameters to represent the interatomic potential. $k$ is the number of parameters to be fit. The potential energy can be written as

$$U = U(\mathbf{R}, \mathbf{X}). \quad (1)$$

The Gibbs free energy difference between two interatomic potentials described by $\mathbf{X}_0$ and $\mathbf{X}$, respectively, can be obtained via HTI [32] as

$$G(\mathbf{X}) - G(\mathbf{X}_0) = \int_{\mathbf{X}_0}^{\mathbf{X}} \left\langle \frac{\partial U}{\partial \mathbf{X}'} \right\rangle_{\mathrm{NPT}, \mathbf{X}'} d\mathbf{X}', \quad (2)$$

where $\mathbf{X}'$ is a point in parameter space along the path from $\mathbf{X}_0$ to $\mathbf{X}$. $\left\langle \frac{\partial U}{\partial \mathbf{X}'} \right\rangle_{\mathrm{NPT}, \mathbf{X}'}$ is the gradient of the potential energy at $\mathbf{X}'$, averaged over the isothermal–isobaric (NPT) ensemble described by the interatomic potential with parameter set $\mathbf{X}'$.

Eq. (2) is exact and allows one to approach a target Gibbs free energy, $G_{target}$, from an initial guess $G(\mathbf{X}_0)$, if a proper path from $\mathbf{X}_0 \to \mathbf{X}_{target}$ can be identified. Here, $\mathbf{X}_{target}$ denotes the parameter set defining the interatomic potential that yields $G_{target}$, i.e.,

$$G_{target} = G(\mathbf{X}_0) + \int_{\mathbf{X}_0}^{\mathbf{X}_{target}} \left\langle \frac{\partial U}{\partial \mathbf{X}'} \right\rangle_{\mathrm{NPT}, \mathbf{X}'} d\mathbf{X}'. \quad (3)$$

We find that $\mathbf{X}_{target}$ can be searched numerically via iterative algorithms. To this end, we employ Newton-Raphson method by iterating intermediate parameter sets, $\mathbf{X}_n$, as

$$\mathbf{X}_{n+1} = \mathbf{X}_n - \frac{G(\mathbf{X}_n) - G_{target}}{\nabla G(\mathbf{X}_n)}, \quad (4)$$

where $\nabla G(\mathbf{X}_n)$ is the free energy gradient with respect to $\mathbf{X}$ at $\mathbf{X}_n$. At each iteration, $G(\mathbf{X}_n)$ is expected to approach $G_{target}$ along the direction of $\nabla G(\mathbf{X}_n)$. Based on the thermodynamic relation [32], we find

$$\nabla G(\mathbf{X}_n) = \left\langle \frac{\partial U}{\partial \mathbf{X}} \right\rangle_{\mathrm{NPT}, \mathbf{X}_n}. \quad (5)$$

Then, Eq. (4) can be written as

$$\mathbf{X}_{n+1} = \mathbf{X}_n - \frac{G(\mathbf{X}_n) - G_{target}}{\left\langle \frac{\partial U}{\partial \mathbf{X}} \right\rangle_{\mathrm{NPT}, \mathbf{X}_n}}. \quad (6)$$

Thus, one can iteratively update $\mathbf{X}_n$ until $G(\mathbf{X}_n) = G_{target}$. Note that when $\left\langle \frac{\partial U}{\partial \mathbf{X}} \right\rangle_{\mathrm{NPT}, \mathbf{X}}$ is constant, Eq. (6) reduces to Eq. (3). In that case, $\mathbf{X}_{target}$ can be directly obtained without iterations, which would imply a linear dependence of $U$ on $\mathbf{X}$, a situation that is generally not the case.

For many-body interatomic potentials, $\mathbf{X}$ usually contains many terms. For each $X_k$ term, the ensemble average $\left\langle \frac{\partial U}{\partial x_k} \right\rangle_{X_n}$ should be computed for Eq. (6). This can be done by treating each derivative $\frac{\partial U}{\partial x_k}$ as an independent Hamiltonian acting on the same ensemble $\langle \ldots \rangle_{X_n}$, which is generated by the Hamiltonian $\mathbf{X}_n$. With this approximation, one does not need to re-perform MD simulations, but only needs to re-calculate the energy using the Hamiltonian $\frac{\partial U}{\partial x_k}$ for the atomic configurations in the ensemble $\langle \ldots \rangle_{X_n}$. We call this as RECAL protocol. The ensemble $\langle \ldots \rangle_{X_n}$ can also be replaced by that from AIMD simulation, which in principle should be the most accurate target ensemble. To reduce fitting complexity, one can also choose a limited number of $X_k$ while fixing other terms in $\mathbf{X}_n$. Then only a few $\left\langle \frac{\partial U}{\partial x_k} \right\rangle_{X_n}$ for the $x_k$ of interest need to be calculated.

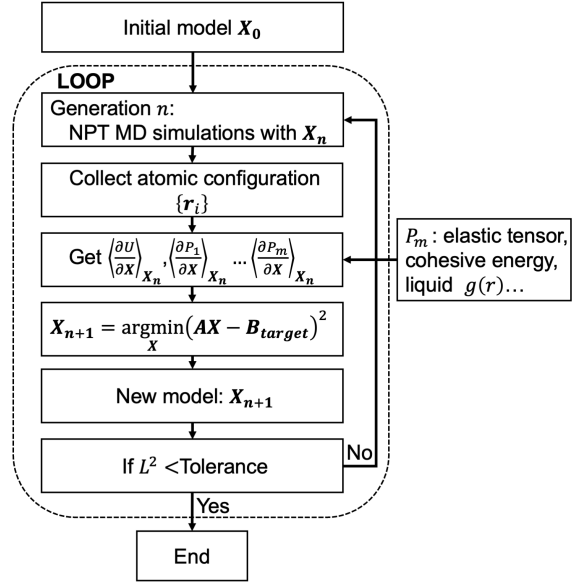

**FIG. 1** Workflow of interatomic potential fitting.

Figure 1 shows the general fitting workflow. Starting from an initial guess of parameter set $\mathbf{X}_0$, an iterative loop is executed, during which NPT-MD simulations are performed to update the parameter set $\mathbf{X}_n$. The energy gradients $\left\langle \frac{\partial U}{\partial \mathbf{X}} \right\rangle_{\mathrm{NPT}, \mathbf{X}_n}$ are computed using atomic configurations $\{\mathbf{r}_i\}$ obtained from the MD trajectory. The model $\mathbf{X}_n$ can be updated to $\mathbf{X}_{n+1}$ via Eq. (6) for free energy fitting. For parameter sets $\mathbf{X}$ containing many parameters, the free energy may not uniquely constrain the parameter set. Therefore, additional high-temperature properties, such as the elastic tensor, liquid structure [19], etc., should also be included as fitting targets to ensure a reasonable solution $\mathbf{X}$. These properties, noted as $P_1, P_2, \ldots, P_m$, are fit together with the free energy $G$ using Newton-Raphson method.

We write down the general fitting equations. Our target set combining the free energy and other properties, is denoted as $(G, P_1, P_2, \ldots, P_m)$. The iteration matrix is

$$A = \left( \left\langle \frac{\partial U}{\partial X} \right\rangle_{X_n}, \left\langle \frac{\partial P_1}{\partial X} \right\rangle_{X_n} \cdots, \left\langle \frac{\partial P_m}{\partial X} \right\rangle_{X_n} \right)^T, \quad (7)$$

where $\left\langle \frac{\partial P_m}{\partial X} \right\rangle_{X_n}$ is the derivative of the property $P_m$ with respect to the parameter set with $X_n$ ensemble. Similar to Eq. (6), we update the parameter set via

$$AX_{n+1} = \begin{pmatrix} G_{target} - G(X_n) \\ \vdots \\ P_{m,target} - P_m(X_n) \end{pmatrix} + AX_n. \quad (8)$$

For Eq. (8), when $k \geq m+1$, one can always find solution for $X_{n+1}$. When $k < m+1$, $X_{n+1}$ can be computed using optimization algorithms, such as linear regression, by minimizing the loss function as $L^2 = (AX - B_{target})^2$, where $B_{target} = \begin{pmatrix} G_{target} - G(X_n) \\ \vdots \\ P_{m,target} - P_m(X_n) \end{pmatrix} + AX_n$. The iteration loop in Fig. 1 ends until $L^2$ is less than a tolerance.

**B. Relative free energy and free energy of mixing**

In practical simulations, the relative free energy difference can be important for predicting phase competition and phase diagram. Thus, it is essential to fit the relative free energy difference among different phases. To achieve this, we can rewrite Eq. (6) for two phases denoted by $L$ and $S$, respectively, as

$$\begin{aligned} G_L(X_n) - G_{L,target} &= -\left\langle \frac{\partial U}{\partial X} \right\rangle_L (X_{n+1} - X_n) \\ G_S(X_n) - G_{S,target} &= -\left\langle \frac{\partial U}{\partial X} \right\rangle_S (X_{n+1} - X_n) \end{aligned}. \quad (9)$$

Subtracting the two equations leads to the relative free energy difference $\Delta G^{L-S} = G_L - G_S$. We can obtain

$$X_{n+1} = X_n - \frac{\Delta G^{L-S}(X_n) - \Delta G^{L-S}_{target}}{\left\langle \frac{\partial U}{\partial X} \right\rangle_L - \left\langle \frac{\partial U}{\partial X} \right\rangle_S}, \quad (10)$$

where $\Delta G^{L-S}_{target}$ is the target free energy difference and $\Delta G^{L-S}(X_n)$ is the free energy difference of the potential with parameter set $X_n$. $\left\langle \frac{\partial U}{\partial X} \right\rangle_L - \left\langle \frac{\partial U}{\partial X} \right\rangle_S$ is the gradient term that needs to be obtained by performing ensemble average for both $L$ and $S$ phases.

When $L$ and $S$ correspond to the liquid and solid phases, one can adjust $X_n$ so that $\Delta G^{L-S}(X_n) = 0$ at $T = T_m$. This is indirectly fitting the melting point $T_m$. One can further compute the temperature-dependent solid-liquid free energy differences using the Gibbs-Helmholtz equation (GHE)

$$\Delta G^{L-S}(X_n, T) = -T \int_{T_m}^{T} \frac{\Delta H^{L-S}(X_n, T)}{T^2} dT. \quad (11)$$

where $\Delta H^{L-S}(X_n, T)$ is the latent heat. Then, $\Delta G^{L-S}(X_n, T)$ can be fitted if a target value is available.

For multicomponent systems, the free energy of mixing is important in determining their phase diagram. The composition-dependent free energy curve provides the fundamental connection to convex hull analysis and chemical potential. When components $A$ and $B$ are mixed to form $A_{1-x}B_x$ in a solution phase $\varphi$ at constant temperature and pressure, the Gibbs free-energy of mixing is

$$G^{\varphi}_{mix} = G^{\varphi}_{A_{1-x}B_x} - xG^{\varphi}_B - (1-x)G^{\varphi}_A. \quad (12)$$

According to Eq. (6), the Gibbs free energies $G^{\varphi}_{A_{1-x}B_x}$ can be expressed as

$$G^{\varphi}_{A_{1-x}B_x}(X_n) - G^{\varphi}_{A_{1-x}B_x,target} = -\left\langle \frac{\partial U}{\partial X} \right\rangle^{\varphi}_{A_{1-x}B_x, X_n} (X_{n+1} - X_n) \quad (13)$$

It can yield a fitting equation for a target mixing free energy $G^{\varphi}_{mix,target}$ as

$$G^{\varphi}_{mix}(X_n) - G^{\varphi}_{mix,target} = -\left\langle \frac{\partial U}{\partial X} \right\rangle^{\varphi}_{mix, X_n} (X_{n+1} - X_n), \quad (14)$$

where $\left\langle \frac{\partial U}{\partial X} \right\rangle^{\varphi}_{mix, X_n} = \left\langle \frac{\partial U}{\partial X} \right\rangle^{\varphi}_{A_{1-x}B_x, X_n} - (1-x)\left\langle \frac{\partial U}{\partial X} \right\rangle^{\varphi}_{A, X_n} - x\left\langle \frac{\partial U}{\partial X} \right\rangle^{\varphi}_{B, X_n}$, which can be calculated based on ensembles of A, B and $A_{1-x}B_x$ systems.

**C. Fitting algorithm for EAM potential**

In this work, we apply our free energy fitting approach on the embedded atom method (EAM) potentials [7,8], which are widely used in the study of metallic systems. In the EAM, the potential energy is composed of a pairwise term $\varphi$ and an embedding term $F$ as

$$U = \sum_i^N \sum_{j=i+1}^N \varphi(r_{ij}) + \sum_i^N F(\rho_i), \quad (15)$$

where $N$ is the number of atoms in the system, $r_{ij}$ is the distance between atoms $i$ and $j$, and $\rho_i$ is the charge density on atom $i$. The function $F$ represents the embedding energy as a function of the local charge density. The functions $\varphi, F$ and $\psi$ need to be specified to fully define the EAM potential. Here, we employ a form of polynomial functions [19] to express $\varphi, F$, and $\rho$ as

$$\varphi(r) = \sum_{k=1}^{n_a} a_k (r_k - r)^{p_k} H(r_k - r) H(r - r_c), \quad (16)$$
$$F(\rho) = -\sqrt{\rho} + \sum_{k=1}^{n_b} b_k (\rho - \rho_k)^{q_k} H(\rho - \rho_k), \quad (17)$$
$$\rho = c(r - r_c)^d \exp(-kr) H(r - r_c), \quad (18)$$

where $H(x)$ is the Heaviside step function. The terms $(r_k - r)^{p_k} H(r_k - r) H(r_c - r)$ and $(\rho - \rho_k)^{q_k} H(\rho - \rho_k)$ are basis functions. $n_a$ and $n_b$ are the number of basis functions in the pairwise and embedding terms, respectively. The parameters in the charge density term $\rho$ are fixed in the free energy fitting. Thus, the adjustable parameters are the weights of the basis function in the pairwise and embedding terms. We can write the parameter set $X$ as

$$X = (a_1, a_2 \cdots a_k, b_1 \cdots b_k). \quad (19)$$

By defining the set of basis functions as

$$\boldsymbol{\Phi} = (\varphi_1, \varphi_2 \cdots \varphi_k, F_1 \cdots F_k)^T, \quad (20)$$

where $\varphi_k = (r_k - r)^{p_k} H(r_k - r) H(r_c - r)$ and $F_k = (\rho - \rho_k)^{p_k} H(\rho - \rho_k)$, the total potential energy of the system can be written as a linear combination of contributions from each basis function

$$U = \boldsymbol{X}\boldsymbol{\Phi}. \quad (21)$$

These functions make it convenient to derive the gradient required by Eq. (6). The derivative with respect to parameters in the pairwise term is

$$\left\langle \frac{\partial U}{\partial a_k} \right\rangle_X = \langle (r_k - r)^{p_k} H(r_k - r) H(r_c - r) \rangle_X = \langle \varphi_k \rangle_X. \quad (22)$$

For the embedding term, the derivative is

$$\left\langle \frac{\partial U}{\partial b_k} \right\rangle_X = \langle (\rho - \rho_k)^{q_k} H(\rho - \rho_k) \rangle_X = \langle F_k \rangle_X. \quad (23)$$

$\langle \ldots \rangle_X$ indicates the ensemble average of the configurations generated by the potential with parameter set $\boldsymbol{X}$. Thus, the derivative is equal to the averaged energy of each basis function.

### D. Simulation details

The MD simulations were performed using the Large-scale Atomic/Molecular Massively Parallel Simulator (LAMMPS) code [33]. In the constant number of atoms, volume, and temperature (NVT) simulation, the Nosé-Hoover thermostat [34,35] was applied. In NPT simulations, the Nosé-Hoover thermostat and barostat were applied. The damping time in the Nose-Hoover thermostat was set to $\tau$=0.01 $ps$. The time step of the simulation was 1.0 $fs$. The melting temperature is obtained via solid–liquid coexistence (SLC) simulations [36]. The elastic tensor was computed by explicit deformation method [37]. The free energy of the binary liquid phase was computed by the non-eqilibrium TI method [33,38]. The target *ab initio* data were obtained from previous studies [29,39,40]. The details of density functional theory calculation to generate these data are provided in Supplemental Material Text S1 [41].

### III. RESULTS
### A. The Uhlenbeck-Ford model

We first demonstrate the fitting approach by adjusting the free energy of a toy model, the Uhlenbeck-Ford model (UFM) [42]. The UFM describes particles interacting via

$$U_{UF}(\sigma) = -\frac{p}{\beta} ln\left[1 - e^{-\left(\frac{r}{\sigma}\right)^2}\right], \quad (24)$$

where $\beta = \frac{1}{k_B T}$. $p$ is the energy-scale parameter and $\sigma$ is the length-scale parameter. The model is often used as a reference state for calculating the free energy of liquid systems. It has an analytical form for its free energy [42], as

$$F = F^{ideal}(\rho) + \frac{1}{\beta} \sum_{n=1}^{\infty} \frac{B_{n+1}(p)}{n} x^n, \quad (25)$$

where $x = \frac{1}{2}(\pi\sigma^2)^{3/2}\rho$. $B_{n+1}$ is the reduced virial coefficient which depends only on the scaling factor $p$ and can be computed exactly [42]. $F^{ideal}(\rho)$ is the contribution of the ideal gas. We set $\rho, \beta, p = 1$ so that only parameter $\sigma$ affects the model.

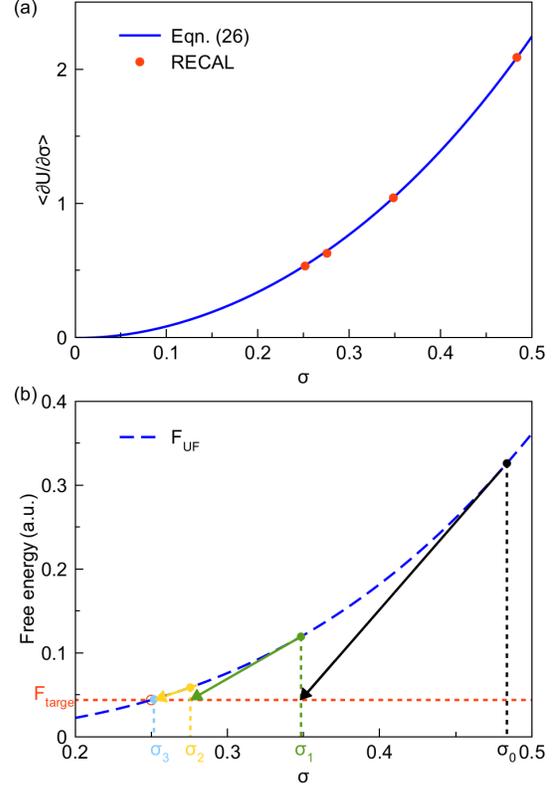

**FIG. 2** Test of the free energy fitting procedure with the UFM model. (a) The ensemble average $\left\langle \frac{\partial U}{\partial \sigma} \right\rangle$ from Eq. (26) and numerical scheme of RECAL. (b) The iterative process starting from an initial guess $\sigma_0$, following the gradient evaluated by $\left\langle \frac{\partial U}{\partial \sigma_0} \right\rangle$. The blue dashed curve represents the free energy profile. The solid line shows the slope $\left\langle \frac{\partial U(\sigma_n)}{\partial \sigma} \right\rangle$ at each iteration. The red horizontal dotted line marks the target free energy. Vertical dotted lines indicate the change of $\sigma$ during iteration.

We set a $F_{target}$ as the target free energy value and $\sigma_0$ as an initial parameter value. The analytical derivative of the free energy with respect to $\sigma$ is

$$\frac{\partial F}{\partial \sigma} = \frac{3}{2} \pi^{\frac{3}{2}} \sigma^2 \frac{1}{\beta} \sum_{n=1}^{\infty} B_{n+1}(p) x^{n-1}. \quad (26)$$

We now use Eq. (6) to iteratively update $\sigma$ so that $F$ approaches $F_{target}$. In this demonstration, the ensemble-averaged derivative $\left\langle \frac{\partial U}{\partial \sigma} \right\rangle$ is evaluated numerically with the RECAL protocol described in Section II.A. Figure 2(a) shows that the gradient obtained from RECAL is consistent with the analytical result from Eq. (26). In Fig. 2(b), we start from an initial guess $\sigma_0$. By updating $\sigma$ with Eq. (6), $\sigma_0$ changes sequentially to $\sigma_1$, $\sigma_2$ and $\sigma_3$. At the third iterations, the $\sigma_3$ is found, yielding a free energy very close to the target value $F_{target}$. This demonstrates the effectiveness of the free energy fitting method.

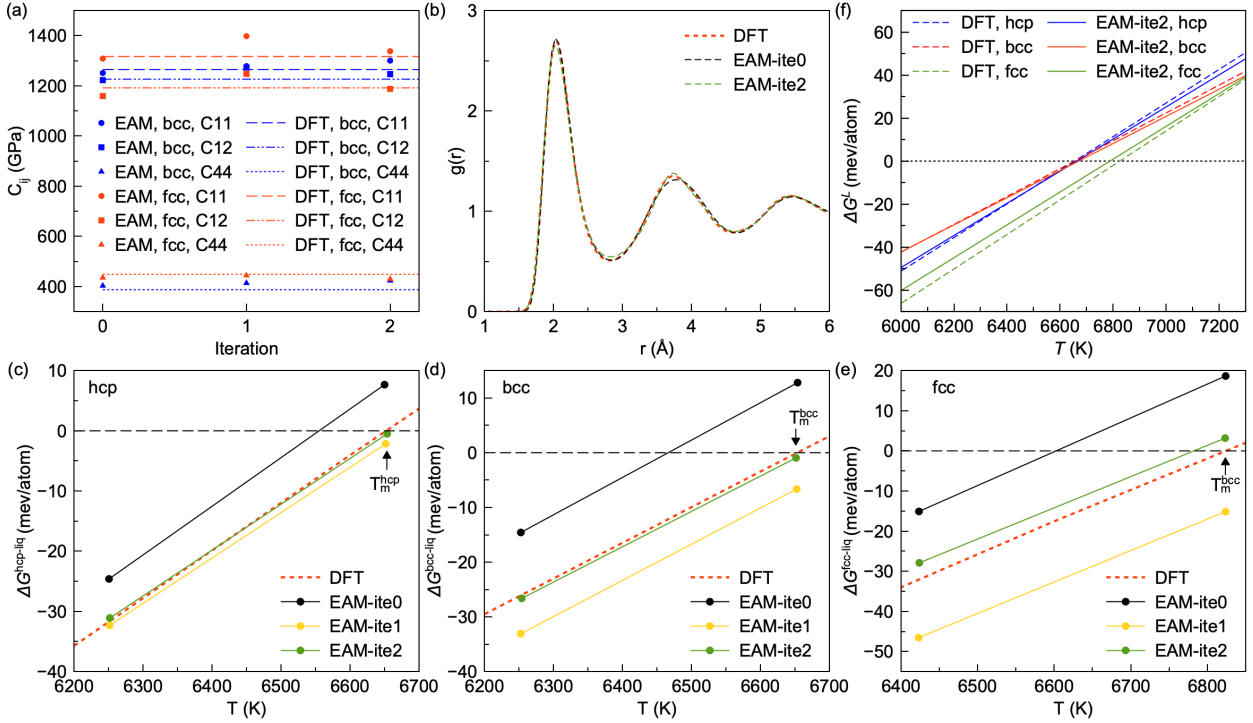

**FIG. 3** Iterative fitting process of the EAM potential for Ni at 323 GPa. (a) Elastic constants of bcc and fcc Ni. Symbols represent EAM values at each iteration; dashed lines denote corresponding DFT values from Ref. [40]. (b) Comparison of liquid radial distribution function g(r) between DFT and EAM potentials. (c–e) Free energy differences between liquid and (c) bcc, (d) hcp, and (e) fcc Ni across iterations. Red dashed lines indicate *ab initio* values from Ref. [29]. The free energy data at $T_m$ and $T_m$–400 K serve as the fitting targets. The resulting free energy values (dots) are connected by lines for clarity. (f) Comparison of solid–liquid free energy differences between DFT and the final EAM potential for all three phases, plotted on the same scale.

### B. EAM potential of nickel system

Next, we apply our fitting approach to an EAM potential. Here, the target is to develop an EAM potential for Ni under extremely high pressure and temperature conditions. In previous work, the *ab initio* free energy difference between liquid and solid Ni phases was computed at 323 GPa and 6000-7000 K, using high-accuracy DFT settings [29]. To fit these data, we employ the EAM potential with the formulas described by Eq. (16-18). It contains $n_a = 20$ pairwise terms with the power $p_k$ ranging from 4 to 8 and $r_k$ value of 2.8, 3.8, 4.8 and 6 Å. The number of embedding terms is $n_b = 9$ with the power $q_k$ ranging from 4 to 6 and $\rho_k$ of 50, 60 and 70. In total, the parameter set $X$ contains 29 fitting parameters. The charge density term $\rho$ is obtained by scaling that of Fe's EAM potential from Ref. [28] based on the ratio of the electron numbers between Fe and Ni.

Several *ab initio* datasets are included in the potential development. First, the *ab initio* free energy data is taken from Ref. [29]. The target values include $\Delta G^{S-L}(T)$ curve ranging from 6200K to 7000K for bcc, hcp, and fcc phases at 323GPa. Complementing these targets, the liquid pair distribution function g(r) and the elastic constants of the bcc and fcc phases are also included, both computed from AIMD simulations [40]. During the fitting process, the melting temperature $T_m$ was determined with SLC simulations using supercells containing 65,520, 64,000 and 65,536 atoms for liquid-hcp, liquid-bcc, and liquid-fcc coexistence structures, respectively. The latent heat was obtained from independent NPT simulations with 7040, 6750, 6912 and 6750 atoms for the hcp, bcc, fcc and liquid, respectively.

The initial EAM potential was fitted to liquid g(r) and elastic constants of bcc and fcc phases based on the atomic configurations obtained from AIMD simulations, noted as the ensemble $X_0$. In this iteration (ite0), the elastic constants and liquid g(r) are well fitted, as shown in Fig. 3(a) and (b), respectively. However, the free energy of hcp, fcc, and bcc computed with the ite0 potential deviates from the target value by 10-20 meV/atom. This results in the melting points deviating by more than ~200 K for the fcc phase, as shown in Fig. 3(c)-(e).

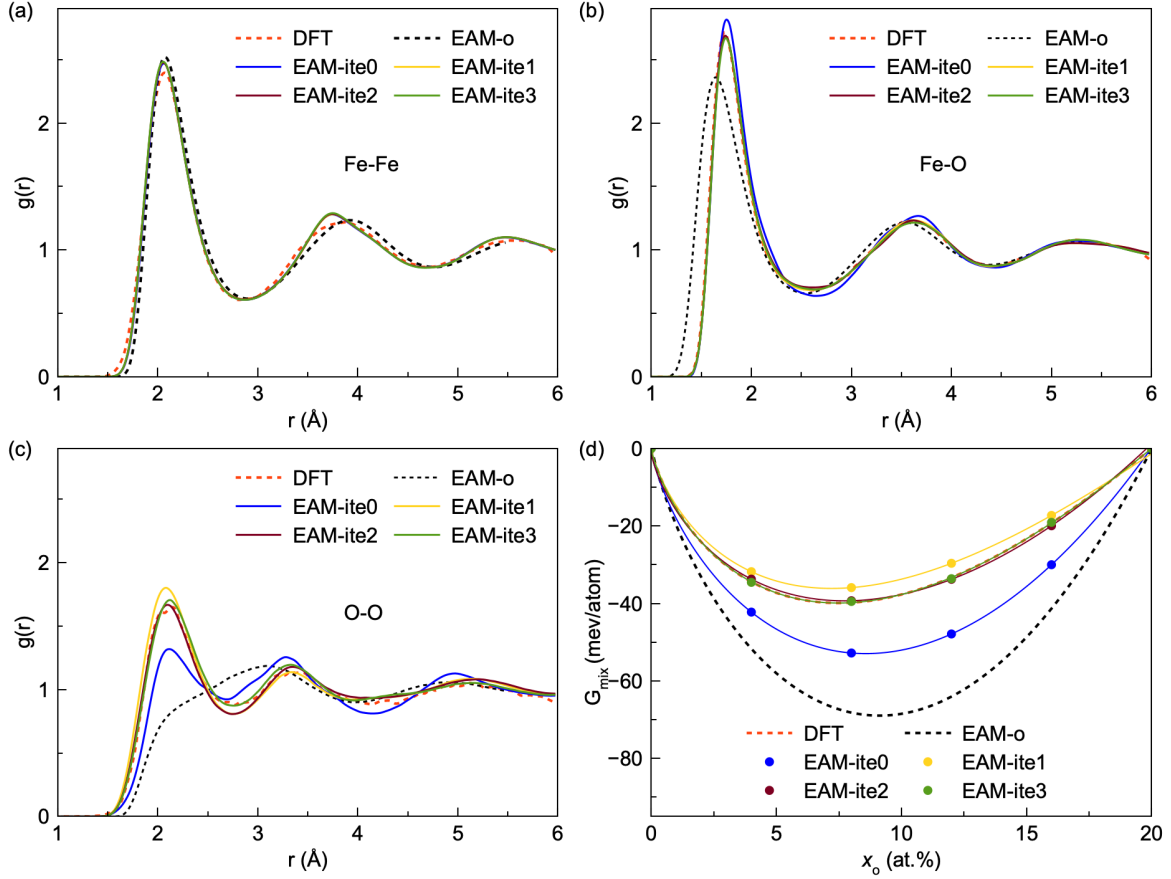

**FIG. 4** Iterative fitting process of the EAM potential for Fe$_{1-x}$O$_x$ liquid at 323 GPa and 5500 K. Evolution of the Fe$_{80}$O$_{20}$ liquid $g(r)$ for (a) Fe-Fe, (b) Fe-O, and (c) O-O pairs. (d) The mixing free energy of Fe$_{1-x}$O$_x$ liquid. Red and black lines are the DFT *ab initio* data and original EAM potential (EAM-o) data from Ref. [39].

Next, the relative free energy differences with respect to liquid phase for hcp, fcc, and bcc are included in the fitting, together with the elastic constants and the liquid $g(r)$. Based on the EAM formulation in Eq. (16)-(18), the parameters $c, d, r_c, k$ in $X$ are fixed, while the parameters $a_k, b_k$ are varied during the fitting process in the following iterations. The derivative of the internal energy has an analytical form according to Eq. (22)-(23), which allows the terms in Eq. (6) to be solved. The fitting was carried out via the iteration scheme described in Eq. (8). For each solid phase, we select *ab initio* free energy data at two temperatures as the fitting targets: one at the melting temperature and one at 400 K below the melting temperature, as shown by the dots in Fig. 3(c)-(e). We monitor the fitting process using the root mean squared error (RMSE), defined as $RMSE = \sqrt{\frac{1}{N}\sum_{i=1}^{N}(P_i - P_{i,target})^2}$, where $N$ is the number of data points of the $P$ property. As shown in Supplemental Material Fig. S1(a) [41], the RMSE of $g(r)$ remains below 0.05 throughout the iterations, whereas the RMSE of the free energy targets decreases systematically. We adopt a free-energy RMSE of 3 meV/atom as the stopping criterion for the iterative fitting procedure here. We find that the free energy values from the EAM potentials converge to the target DFT value within only two iterations. The evolution of elastic constants for the bcc and fcc phases during the fitting is monitored in Fig. 3(a), where only minor deviations from the DFT targets were observed. Figure 3(b) demonstrates that the final EAM-ite2 potential accurately reproduces the *ab initio* liquid $g(r)$. Figure 3(f) shows that the Gibbs free energy differences w.r.t. liquid for hcp, bcc, and fcc phases described by the EAM-ite2 potential agree well with DFT results, with RMSE of Gibbs free energy of 2.5 meV/atom. This yields a melting temperature for the fcc phase within 40 K of the DFT target. These results indicate that the present scheme is effective for fitting high-temperature free energy data of Ni phases.

## C. EAM potential of iron-oxygen liquid

We further test the effectiveness of the method by fitting the *ab initio* mixing free energy of the binary $Fe_{1-x}O_x$ liquid under high pressure and high temperature conditions. Previous studies reported the *ab initio* mixing free energy of $Fe_{1-x}O_x$ solutions for $x$ ranging from 0 to 20% under 323 GPa and 5500 K [39]. We test the current method by fitting the relative mixing free energy using $G(x = 0)$ and $G(x = 20\%)$ as the reference states.

The binary EAM potential is developed in the Finnis-Sinclair (FS) form [8], which includes three components: Fe-Fe, Fe-O and O-O interactions. The Fe-Fe interaction is fitted by repeating the process described for Ni system in Section III.B. This EAM potential accurately reproduces the *ab initio* free energies of the hcp and liquid phases of pure iron reported in Ref. [27]. We then fix the parameters of pairwise, embedding and density terms for Fe-Fe interactions. For O-O, we set the density term by scaling Fe's density term according to the ratio of the electron numbers of O and Fe. This approximation is useful here as the iron-oxygen phases is metallic under such high-pressure conditions, and the potential is intended for systems with low oxygen concentrations. The Fe-O and O-O interactions are then fitted by adjusting the parameters in their pairwise terms, as well as the embedding term of O. The original EAM (EAM-o) potential from Ref. [39] is used to generate the initial ensemble. As shown in Fig. 4(b)-(d), both liquid $g(r)$ and the mixing free energy described by the EAM-o potential deviate significantly from the *ab initio* results. After incorporating the *ab initio* liquid $g(r)$ of $Fe_{80}O_{20}$ and the mixing free energy of liquid phases at oxygen concentrations of 4%, 8%, 12% and 16%, the optimized EAM potential reproduces these properties within only three iterations. The convergence criterion is set to a mixing free-energy RMSE of 1 meV/atom. The final EAM-ite3 potential provides an excellent match to the *ab initio* free energies and pair distribution functions of the liquid, as shown in Fig. 4. The final mixing free-energy RMSE is only 0.4 meV/atom, as shown in Supplemental Material Fig. S1(b) [41].

To examine the effect of the initial conditions on the fitting convergence, we performed a robustness test by introducing small variations in the parameter set to construct four different initial potentials. These initial models yield noticeably different free-energy predictions, as shown in Supplemental Material Fig. S2 [41], with the initial RMSE of the mixing free energy spanning approximately 2–27 meV/atom. After one fitting iteration, the errors in all four cases fall below the convergence criterion (1 meV/atom) and remain stable over the subsequent two iterations. The liquid g(r) also converges well. These tests indicate that the method is robust with respect to variations in the initial parameter set.

Regarding the fitting efficiency, we note that the actual parameter-fitting time in our scheme is negligible. The dominant computational cost arises from the multiple intermediate simulations required to obtain the properties of the fitted potentials. Nevertheless, the overall fitting cost remains much lower than that of training a machine-learning potential on a large dataset of atomic energies and forces (see Supplemental Material Text 2 [41]).

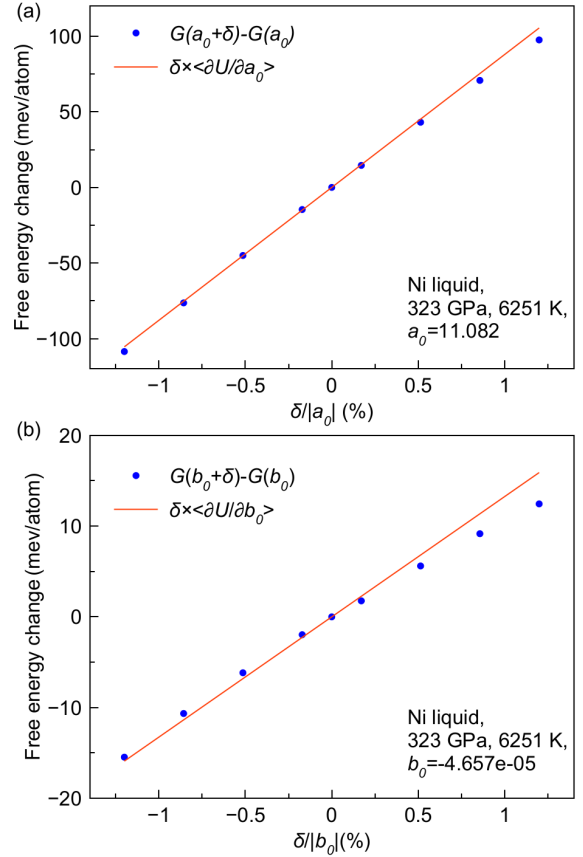

**FIG. 5** Comparison between the gradient method and the perturbation method for Ni liquid at 323 GPa and 6251 K. (a) Free energy change due to a perturbed pair parameter $a_0$. Blue symbols show the free energy change obtained using HTI. The red line represents the analytical result from the internal energy gradient. (b) Free energy change due to a perturbed embedding parameter $b_0$.

## IV. DISCUSSION

Using the Ni and $Fe_{1-x}O_x$ systems, we show that our method can accurately fit the Gibbs free energy during the development of the EAM potential. The demonstrations applied to both UFM and EAM potentials validate the proposed methodology. The key idea of the method is to use the derivative of the internal energy with respect to the parameter set, i.e., Eq. (6), to guide the free energy fitting. These easily accessible quantities, together with derivatives of other desired properties with respect to the parameters, make the fitting process controllable. The EAM potentials used here employ polynomial functions, which provide an analytical way of computing the derivatives for free energy and other properties. Thus, the parameter fitting can be carried out using a Newton–Raphson scheme without involving any step-size parameter. This, however, is not a prerequisite for the fitting method. In principle, one can always use a perturbation approach to compute the derivative $\frac{\partial P_m}{\partial X}$ in Eq. (7). For a specific parameter $x_k$, the derivative can be obtained by applying a sufficiently small perturbation $\delta$ as

$$\frac{\partial P_m}{\partial X_k} \approx \left\langle \frac{P_m(X_k+\delta) - P_m(X_k)}{\delta} \right\rangle_{X_n}. \quad (27)$$

Based on the developed Ni EAM potential, we provide a simple test of the perturbation method on the free energy. In Fig. 5(a) and (b), we independently perturb a parameter in a pairwise term ($a_0$), and a parameter in the embedding term ($b_0$), respectively. The resulting change in liquid free energy $G(a_0 + \delta) - G(a_0)$ and $G(b_0 + \delta) - G(b_0)$ are evaluated by the HTI in Eq. (2), and compared with the results of the analytical expression $\delta \times \left\langle \frac{\partial U}{\partial X} \right\rangle$. The results show that the free energy changes from the perturbation are consistent with the analytical value, with discrepancies of less than 1 meV/atom, provided the perturbation is smaller than 0.8% for the pairwise term and 0.5% for the embedding parameter. Therefore, for potentials without an explicit formula for the free energy gradient, one can still fine-tune the free energy using the perturbation method, albeit at a higher computational cost. Thus, in future work, our free energy fitting method can be further applied to machine learning potentials, which are currently trained primarily on the energies and forces of atomic configurations.

## V. CONCLUSION

We have developed a method for fitting high-temperature Gibbs free energy in the development of interatomic potentials. By reformulating the HTI equations, we derived an iterative scheme that guides potential parameter adjustment based on ensemble-averaged energy derivatives. We demonstrated the accuracy and versatility of this method through three representative systems. First, a toy model based on the Uhlenbeck–Ford potential confirmed the stability and rapid convergence of the scheme. Then, we applied the method to fit an EAM potential for nickel under high pressure conditions, using *ab initio* free energy data. The resulting EAM potential reproduces solid-liquid free energy differences and melting temperatures within a few iterations, achieving satisfactory agreement with DFT data. We further extended the framework to the binary Fe-O liquid system, successfully fitting the composition-dependent mixing free energy and liquid structure. These applications demonstrate the method's ability to incorporate both unary and binary thermodynamic data. While demonstrated here for EAM potentials, the method is general and can be applied to other potential forms.

## ACKNOWLEDGMENTS

Work at Xiamen University was supported by the National Natural Science Foundation of China (Grants Nos. T2422016 and 42374108). S. Fang and T. Wu from the Information and Network Center of Xiamen University are acknowledged for their help with Graphics Processing Unit computing. The supercomputing time was partly supported by the Opening Project of the Joint Laboratory for Planetary Science and Supercomputing, Research Center for Planetary Science, and the National Supercomputing Center in Chengdu (No. CSYYGS-QT-2024-15).